\documentstyle[epsf,prd,aps,floats,epsfig]{revtex}

\begin{document}

\title{Solving the Flatness and Quasi-flatness Problems in Brans-Dicke Cosmologies
with a Varying Light Speed}
\author{John D. Barrow$^1$ and Jo\~ao Magueijo$^2$}
\address{$^1$Astronomy Centre, University of Sussex, Brighton BN1~9QJ, U.K.\\
$^2$ Theoretical Physics, The Blackett Laboratory,\\
Imperial College, Prince Consort Road, London, SW7~2BZ, U.K.}
\maketitle

\begin{abstract}
We define the flatness and quasi-flatness problems in cosmological models.
We seek solutions to both problems in homogeneous and isotropic Brans-Dicke
cosmologies with varying speed of light. We formulate this theory and find
perturbative, non-perturbative, and asymptotic solutions using both
numerical and analytical methods. For a particular range of variations of
the speed of light the flatness problem can be solved. Under other
conditions there exists a late-time attractor with a constant value of $
\Omega \ $that is smaller than, but of order, unity. Thus these theories may
solve the quasi-flatness problem, a considerably more challenging problem
than the flatness problem. We also discuss the related $\Lambda $ and quasi-$
\Lambda $ problem in these theories. We conclude with an appraisal of the
difficulties these theories may face.
\end{abstract}

\date{\today }


\renewcommand{\thefootnote}{\arabic{footnote}} \setcounter{footnote}{0}

\section{The flatness and the quasi-flatness problems}

The observable universe is close to a flat Friedmann model, the so-called
Einstein-de Sitter universe, in which the energy density, $\rho ,$ takes the
critical value, $\rho _c$, and the homogeneous spatial surfaces are
Euclidean. All astronomical evidence shows that we are quite close to this
state of flatness, although a value of $\Omega _0$ in the vicinity of $0.2$
is preferred by several observations.

It is therefore disquieting to notice that the flat Friedmann model
containing dust or blackbody radiation is unstable as time increases. Small
deviations from the exact $\Omega =1$ model grow quickly in time, typically
like $a^2$, where $a$ is the expansion factor of the universe. The observed $
\Omega _0\approx 1$ state therefore requires extreme fine tuning of the
cosmological initial conditions, assumed to be set at Planck epoch, because $
a$ has increased by a factor of order $10^{32}$ from that epoch to the
present. This is the {\it flatness problem}. If the universe is slightly
open at Planck time, within a few Planck times it would become totally
curvature dominated. If it is initially slightly closed, it would quickly
collapse to Planck density again. Explaining its current state requires an
extraordinarily close proximity to perfect flatness initially, or some
sequence of events which can subsequently reverse expectations and render
the flat solution asymptotically stable.

Particle physics theories naturally contain self-interacting scalar matter
fields which violate the strong energy condition (so the density and
pressure, $p$, obey $\rho +3p<0$). These can make the flat solution
asymptotically stable with increasing time, allowing the asymptotic state to
naturally be close to flatness. Cosmological histories in which a brief
period of expansion is dominated by a matter field or other effective stress
which violate the strong energy condition, and so exhibit gravitational
repulsion, are called ``inflationary''.

Solutions to the flatness problem have been proposed in the context of
inflationary scenarios \cite{infl}, pre-Big-Bang models \cite{prebb}, and
varying speed of light cosmologies \cite{mof,mof1,vsl0,vsl1}. In all of
these theories, $\Omega =1$ becomes an asymptotically stable attractor for
the expanding universe. The observed $\Omega _0\approx 1$ state results then
from a temporary period of calculable physical processes, rather than from
highly tuned initial conditions. In such scenarios the price to be paid is
that $\Omega _0$ should be very close to unity, $\left| \Omega _0-1\right| $ 
$\leq 10^{-5}$, if one is not to invoke an unmotivated fine tuning of the
initial conditions again.

If we take the trend of the observational data seriously, then explaining a
current value of $\Omega _0$ of, say, $0.2-0.9$ is yet another challenge. We
call it the {\it quasi-flatness problem}. Solutions to this problem have
been proposed in the context of open inflationary models \cite{open}. In
these one has to come to grips with some degree of fine tuning. The
Anthropic Principle \cite{anth0} is usually invoked for this purpose \cite
{anth} but considerable uncertainties exist in the range of predictions that
emerge and there does not appear to be scope for a very precise prediction
of $\Omega _0$ in, say, the range $0.2.$ Undoubtedly, it would be better if
one could find mechanisms which would produce a definite $\Omega _0$ of
order one, but different from 1, as an attractor. In a recent letter \cite
{quasi}, we displayed one theory in which this possible. Here we present
further solutions in support of such models. We explore analytical and
numerical solutions to Brans-Dicke (BD) cosmologies with a varying speed of
light (VSL). These generalise our earlier investigations of this theory \cite
{vsl0,vsl1,vsl2}. We show that if the speed of light evolves as $
c\propto a^n$ with $0<n<-1$ there is a late-time attractor at $\Omega =-n$.
Hence, these cosmologies can solve the quasi-flatness problem. This work
expands considerably the set of solutions presented in ref. \cite{quasi}.

We note that the existence of the dimensionless fine structure constants
allows these varying-$c$ theories to be transformed, by a change of units,
into theories with constant $c,$ but with a varying electron change, $e$, or
dielectric 'constant' of the vacuum. This process is described in detail in
ref. \cite{vsl2} where a particular theory is derived from an action
principle. A different varying-$e$ theory has been formulated by Bekenstein 
\cite{bek} but it is explicitly constructed to produce no changes in
cosmological evolution. A study of this theory will be given elsewhere \cite
{btoo}.

There also exist analogues of the flatness and quasi-flatness problems with
regard to the cosmological constant. The {\it lambda problem} is to
understand why the cosmological constant term, $\Lambda c^2$ in the
Friedmann equation does not overwhelmingly dominate the density and
curvature terms today (quantum gravity theories suggest that it ought to be
about $10^{120}$ times bigger than observations permit \cite{hawk,anth0}).
The {\it quasi-lambda problem} is to understand how it could be that this
contribution to the Friedmann equation could be non-zero and of the same
order as the density or curvature terms (as some recent supernovae
observations suggest \cite{super}).

In Section~\ref{eqns} we write down the evolution equations in these
theories. The varying $G$ aspect of the theory will be accommodated in the
standard way by means of a Brans-Dicke theory of gravitation. This theory is
adapted in a well-defined way to incorporate varying $c.$ We write equations
in both the Jordan and Einstein's frames. In Section~\ref{pert} we study
solutions to the flatness problem in the perturbative regime ($\Omega
\approx 1$), when both $c$ and $G$ may change. In Section~\ref{gcons} we
present non-perturbative solutions when $G$ is constant, and in Section~\ref
{gvar} when $G$ is varying. In Section \ref{asym} we derive simple
conditions on any power-law variation of $c$ with scale factor if the
flatness problem is to be solved. These conclusions are reinforced by some
exact solutions in Section~\ref{exact}. In section \ref{qflat} we show that
the quasi-flatness problem can be naturally solved in a class of varying-$c$
cosmologies and then we discuss the solution of the quasi-lambda problem by
such cosmologies in Section \ref{qlamb}. We conclude with a summary of our
results, highlighting the cases in which we can claim to have solved the
quasi-flatness and quasi-lambda problems.

\section{Cosmological Field Equations}

\label{eqns} In BD varying speed of light (VSL) theories the Friedmann
equations are \cite{vsl1}: 
\begin{eqnarray}
3{\frac{\ddot a}a} &=&-{\frac{8\pi }{(3+2\omega )\phi }}[(2+\omega )\rho
+3(1+\omega )p/c^2]-\omega {\left( \frac{\dot \phi }\phi \right) }^2-{\frac{
\ddot \phi }\phi }  \label{fr1} \\
{\left( \frac{\dot a}a\right) }^2 &=&{\frac{8\pi \rho }{3\phi }}-{\frac{Kc^2
}{a^2}}-{\frac{\dot \phi a}{\phi a}}+{\frac \omega 6}{\left( \frac{\dot \phi 
}\phi \right) }^2  \label{fr2}
\end{eqnarray}
where the speed of light, $c,$ is now an arbitrary function of time, $K$ is
the curvature constant, and $\omega $ is the constant BD parameter. The wave
equation for the BD scalar field $\phi =1/G$ is 
\begin{equation}
\ddot \phi +3{\frac{\dot a}a}\dot \phi ={\frac{8\pi }{3+2\omega }}(\rho
-3p/c^2)  \label{fr3}
\end{equation}
From these three equations we can obtain the generalised conservation
equation. Since after we impose the equation of state, the varying $c$ term
only appears in eq. (\ref{fr2}), the only new contribution to $2\dot a\ddot
a $ is from the $-2Kc\dot c$ term. Hence, together, these imply the
'non-conservation' equation: 
\begin{equation}
\dot \rho +3{\frac{\dot a}a}(\rho +p/c^2)={\frac{3Kc\dot c}{4\pi a^2}}\phi .
\label{noncons}
\end{equation}
In the radiation-dominated epoch ($p=\rho c^2/3$) the general solution for $
\phi $ is 
\begin{equation}
\phi =\phi _0+\alpha {\int {\frac{dt}{a^3}}}  \label{dotG}
\end{equation}
If the integration constant $\alpha =0$ the usual solutions for VSL in
general relativity with varying $c$ follow illustrating the fact that for a
radiation source any solution of general relativity is a particular solution
of BD theory with constant $\phi $. In the next sections we explore the
solutions to the flatness problem when $\dot \phi \neq 0$. We will also
consider the radiation to matter transition in these theories.

Equations (\ref{fr1})-(\ref{dotG}) apply in the so-called Jordan frame. It
will be useful to introduce the Einstein frame, by means of the
transformations, 
\begin{eqnarray}
d{\hat t} &=&{\sqrt{G\phi }}dt \\
{\hat a} &=&{\sqrt{G\phi }}a \\
\sigma &=&{\left( \omega +3/2\right) }^{1/2}\ln ({G\phi )} \\
{\hat \rho } &=&(G\phi )^{-2}\rho \\
{\hat p} &=&(G\phi )^{-2}p, \\
&&  \nonumber
\end{eqnarray}
which are performed at constant $c$. These may be regarded as merely
mathematical transformations of variables.

The Friedmann equations in the new frame are 
\begin{eqnarray}
3{\frac{{\hat a}^{\prime \prime }}{{\hat a}}} &=&{\frac{-4\pi G}3}[{\hat
\rho }+3{\hat p}/c^2]-{\frac{\sigma ^{\prime }{}^2}3} \\
{\left( \frac{{\hat a}^{\prime }}{{\hat a}}\right) }^2+{\frac{Kc^2}{{\hat a}
^2}} &=&{\frac{8\pi G{\hat \rho }}3}+{\frac{\sigma ^{\prime }{}^2}6}
\end{eqnarray}
where $^{\prime }=d/d{\hat t}$. The transformed scalar field equation for $
\sigma $ is 
\begin{equation}
\sigma ^{\prime \prime }+3{\frac{{\hat a}^{\prime }}{{\hat a}}}\sigma
^{\prime }=-{\frac{8\pi G}{{\sqrt{6+4\omega }}}}({\hat \rho }-3{\hat p}/c^2)
\end{equation}
These are just the standard Friedmann equations with constant $G$ and a
scalar field added to the normal matter. The scalar field behaves like a
'stiff' perfect fluid with equation of state 
\begin{equation}
{\hat p}_\sigma ={\hat \rho }_\sigma ={\frac{\sigma ^{\prime }{}^2}{16\pi G}.
}
\end{equation}
In the Einstein frame, if $\dot c=0$, the total stress-energy tensor is
conserved, but the scalar field and normal matter exchange energy according
to: 
\begin{equation}
{\hat \rho }^{\prime }+3{\frac{{\hat a}^{\prime }}{{\hat a}}}({\hat \rho }+{
\hat p}/c^2)=-({\hat \rho }_\sigma ^{\prime }+3{\frac{{\hat a}^{\prime }}{{
\hat a}}}({\hat \rho }_\sigma +{\hat p}_\sigma /c^2))={\frac{\sigma ^{\prime
}({\hat \rho }-3{\hat p}/c^2)}{{\sqrt{6+4\omega }}}.}
\end{equation}
If $\dot c\neq 0,$ one has instead 
\begin{eqnarray}
{\hat \rho }^{\prime }+3{\frac{{\hat a}^{\prime }}{{\hat a}}}({\hat \rho }+{
\hat p}/c^2) &=&{\frac{\sigma ^{\prime }({\hat \rho }-3{\hat p}/c^2)}{{\sqrt{
6+4\omega }}}}+{\frac{3Kc^2}{4\pi G{\hat a}^2}}{\frac{c^{\prime }}c}
\label{conseins} \\
{\hat \rho }_\sigma ^{\prime }+3{\frac{{\hat a}^{\prime }}{{\hat a}}}({\hat
\rho }_\sigma +{\hat p}_\sigma /c^2) &=&-{\frac{\sigma ^{\prime }({\hat \rho 
}-3{\hat p}/c^2)}{{\sqrt{6+4\omega }}}}
\end{eqnarray}
All equations derived for standard VSL theory, with constant $G$, are valid
in the Einstein frame. However, one should always remember to add to normal
matter the scalar field energy and pressure (so the total density and
pressure are given by ${\hat \rho }_t={\hat \rho }+\rho _\sigma $ and ${\hat
p}_t={\hat p}+{\hat p}_\sigma ,$ respectively).

\section{Perturbative solutions to the flatness problem}

\label{pert} We first study solutions to the flatness problem when there are
small deviations from flatness. Let us define the critical density, $\rho _c$
, in a B-D universe by means of the equation: 
\begin{equation}
{\left( \frac{\dot a}a\right) }^2={\frac{8\pi \rho _c}{3\phi }}-{\frac{\dot
\phi a}{\phi a}}+{\frac \omega 6}{\left( \frac{\dot \phi }\phi \right) }^2,
\end{equation}
that is, Eqn.~\ref{fr2} with $K=0$. In the Einstein frame the critical
density in normal matter is 
\begin{equation}
{\hat \rho }_c={\frac 3{8\pi G}}{\left( {\left( \frac{{\hat a}^{\prime }}{{
\hat a}}\right) }^2-{\frac{{\sigma ^{^{\prime }2}}}6}\right) }={\frac{\rho _c
}{G^2\phi ^2}}
\end{equation}
In terms of total energy density ${\hat \rho }_t={\hat \rho }+\rho _\sigma $
, the critical energy density in the Einstein frame is: 
\begin{equation}
{\hat \rho }_{ct}={\frac 3{8\pi G}}{\left( \frac{{\hat a}^{\prime }}{{\hat a}
}\right) }^2.
\end{equation}
Accordingly, we may define a {\it relative flatness parameter:} 
\begin{equation}
{\hat \epsilon }_t={\frac{{\hat \rho }_t-{\hat \rho }_{ct}}{{\hat \rho }_{ct}
}}
\end{equation}
As shown in the previous section, the usual equations for VSL theory should
apply to this quantity. Therefore, one has 
\begin{equation}
{\hat \epsilon _t}^{\prime }=(1+{\hat \epsilon }_t){\hat \epsilon }
_t(3\gamma -2){\frac{{\hat a}^{\prime }}{{\hat a}}}+2{\frac{c^{\prime }}c}{
\hat \epsilon }_t
\end{equation}
with the equation of state given by

\begin{equation}
\gamma -1\equiv p/\rho c^2={\hat p}/{\hat \rho }c^2  \label{gamma}
\end{equation}
with $\gamma $ constant.

But, since 
\begin{equation}
{\hat \epsilon }_t={\frac \epsilon {1+(2\omega +3){\frac{{\dot \phi }^2}{
8\pi \phi \rho _c}}},}
\end{equation}
we see that the natural quantity to quantify deviations from flatness in the
Jordan frame is not 
\begin{equation}
\epsilon ={\frac{{\rho }-{\rho }_c}{{\rho }_c}}
\end{equation}
but an adaptation of it, 
\begin{equation}
\delta ={\frac{\rho -\rho _c}{\rho _c+{\frac{(2\omega +3){\dot \phi }^2}{
8\pi \phi }}}}<\epsilon ,  \label{deltaeqn}
\end{equation}
which satisfies the equation 
\begin{equation}
\dot \delta =(1+\delta )\delta (3\gamma -2){\left( {\frac{\dot a}a}+{\frac 12
}{\frac{\dot \phi }\phi }\right) }+2{\frac{\dot c}c}\delta .  \label{eps}
\end{equation}
If $\delta \ll 1$ this can be integrated to give 
\begin{equation}
\delta \propto a^{3\gamma -2}c^2\phi ^{\frac{\ 3\gamma -2}2}
\end{equation}
Hence, we have the solution 
\begin{equation}
{\frac 1\epsilon }={\frac C{a^{3\gamma -2}c^2\phi ^{\frac{3\gamma -2}2}}}-{
\frac{(2\omega +3){\dot \phi }^2}{8\pi \phi \rho _c}.}  \label{perteps}
\end{equation}
If $G\propto \phi ^{-1}$ varies then the second term on the right always
works against solving the flatness problem. Hence solving the flatness
problem in BD requires that $\phi $ decreases in the early universe and that
the first term on the right of eq. (\ref{perteps}) dominates the second.

In a radiation-dominated universe ($\gamma =4/3$): 
\begin{equation}
{\frac 1\epsilon }={\frac C{a^2c^2\phi ^2}}-{\frac{(2\omega +3){\alpha }^2}{
8\pi \phi a^6\rho _c},}
\end{equation}
where $\alpha $ was defined by eqn.(\ref{dotG}). If $\alpha $ is small and $
c=c_0a^n$ one still solves the flatness problem when $n<-1$. Notice that
since $\rho _c\propto 1/a^4$ at late times, the second term always
eventually becomes negligible.
 
\section{Non-perturbative solutions with $\phi=const$}

\label{gcons}

\subsection{Matter and radiation-dominated cases}

\begin{figure}[tbp]
\centerline{{\psfig{file=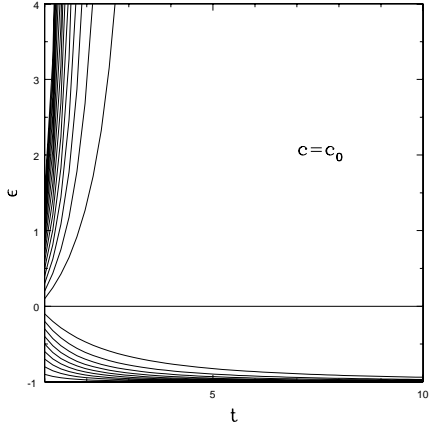,width=6 cm,angle=0}}}
\caption{The evolution of $\epsilon $ from various
initial values for cosmologies with $\phi =const$ and $c=c_0$. We see that
the flat universe is unstable.}
\label{fig1}
\end{figure}

\begin{figure}[tbp]
\centerline{{\psfig{file=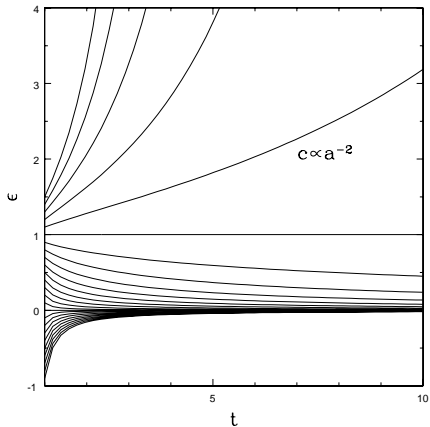,width=6 cm,angle=0}}}
\caption{The VSL solution to the flatness problem.
We have assumed a radiation-dominated universe, and considered a situation
with $n<-(3\gamma -2)/2$. We see that although the flat universe is now an
attractor, if the universe is ever too closed it will end in a Big Crunch,
rather than approach the flat attractor.}
\label{fig2}
\end{figure}

\begin{figure}[tbp]
\centerline{{\psfig{file=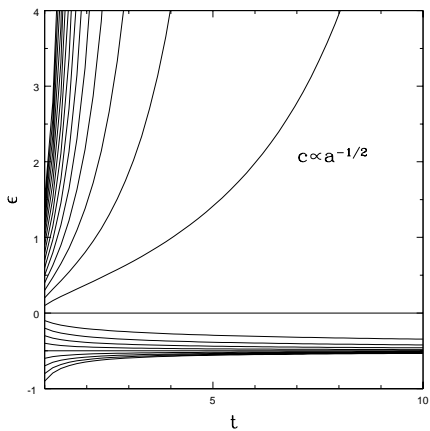,width=6 cm,angle=0}}}
\caption{The VSL solution to the quasi-flatness
problem. We have assumed a radiation-dominated universe, and considered a
situation with $-(3\gamma -2)/2<n<0$. The attractor is now at a value of $
\Omega $ of order 1, but not necessarily 1. Closed universes end up in a Big
Crunch, but any open universes are pushed towards $\Omega =-2n/(3\gamma -2)$.
}
\label{fig3}
\end{figure}

\begin{figure}[tbp]
\centerline{\ {\psfig{file=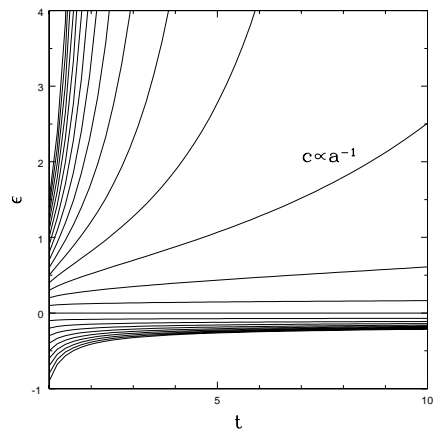,width=6 cm,angle=0}}}
\caption{The transition case $n=-(3\gamma -2)/2$ illustrated for a radiation
dominated Universe.}
\label{fig4}
\end{figure}

In order to explore non-perturbative solutions to the flatness problem we
solved Eqns. (\ref{fr1})-(\ref{dotG}) numerically. First we consider the
case where $\phi =\phi _0$. Exact solutions exist in this case for matter
and radiation-dominated universes. Setting $\phi =1/G$ in Eqn. (\ref{eps})
leads to 
\begin{equation}
{\dot \epsilon }=(1+\epsilon )\epsilon (3\gamma -2){\frac{{\dot a}}{{a}}}+2{
\frac{{\dot c}}c}\epsilon
\end{equation}
which, assuming ${\dot c}/c=n{\dot a}/a,$ becomes 
\begin{equation}
{\dot \epsilon }=\epsilon {\left( (1+\epsilon )(3\gamma -2)+2n\right) }{
\frac{{\dot a}}{{a}}.}
\end{equation}
The general structure of the attractors can be inferred from this equation.
However, it can also be integrated to give 
\begin{equation}
\epsilon =-{\frac{1+{\frac{2n}{3\gamma -2}}}{1-Ca^{-(2n+3\gamma -2)}},}
\label{soleps}
\end{equation}
with $n<0$, and where the integration constant $C$ is chosen so as to
enforce the specified initial conditions. Our numerical code matched this
analytical result to very high accuracy. In Fig.~\ref{fig1} we plot
numerical results for a radiation-dominated universe. We comment on four
possible situations, illustrated in Figs.~\ref{fig1}-\ref{fig4}.

If $c$ is a constant ($n=0$) then the flat universe ($\epsilon =0$) is
unstable (Fig.~\ref{fig1}). There are two attractors at $\epsilon =-1$ and $
\epsilon =\infty $, showing that the universe tends to become curvature
dominated. For slightly closed universes this means evolution towards a Big
Crunch singularity. For slightly open universes it means evolution towards a
Milne universe ($\epsilon =-1$) which in the case of constant $G$ is simply
an expanding empty spacetime with $a\propto t$. If the speed of light were
to increase with $a$ ($c\propto a^n$ with $n>0$) the situation is the same,
but with a stronger repulsion from $\epsilon =0$.

If $c\propto a^n$ with $n<-(3\gamma -2)/2$, then, as pointed out in \cite
{vsl1}, the flatness problem is solved if $\epsilon $ is not too far from
zero initially. In fact, $\epsilon =0$ is now an attractor (see Fig.~\ref
{fig2}). The non-perturbative analysis reveals two novelties: there is an
unstable node at $\epsilon =-1-2n/(3\gamma -2)$ and $\epsilon =\infty $ is
an attractor. This means that if the universe at any given initial time
satisfies $\epsilon >-1-2n/(3\gamma -2)$ it will not evolve towards the flat
attractor. Instead, it will evolve towards a big crunch with $\epsilon
=\infty $. Therefore, for closed universes, the flatness problem is solved
only if they are not too positively curved.

It is curious to note that the situation is different for open universes. No
matter how far they are from flatness they will always evolve towards the
Einstein-de Sitter model at late times. This was first pointed out in \cite
{vsl0}, where it was noted that in VSL theories the Einstein-de Sitter
universe is an attractor even if the universe is initially Minkowski (empty)
or de Sitter (dominated by a cosmological constant).

The greatest novelty, however, appears for the case in which $-(3\gamma
-2)/2<n<0$. Although $\epsilon =0$ is now unstable, there is a stable
attractor at $\epsilon =-1-2n/(3\gamma -2)$ (see Fig.~\ref{fig3}). The
system is now repelled from $\epsilon =-1$ but is attracted to $\epsilon
=\infty $.

Hence we have the following scenario: all closed universes evolve towards a
big crunch, no matter how close to flatness they are initially. Thus a
selection process excludes even slightly closed universes. {\em \ }In
contrast, flat and all open models (no matter how low their density) evolve
towards an open model with $\Omega =-2n/(3\gamma -2)$, a value that lies
between $0$ and $1$, and is typically of order unity. We therefore have a
solution to the quasi-flatness problem, for open, but not for closed
universes.

The $n=-(3\gamma -2)/2$ case is a transition case in between the last two
situations just described (Fig.~\ref{fig4}). In this case $\epsilon =0$ is a
saddle: stable if approached from below but unstable from above.

\subsection{The radiation-to-matter transition}

In quasi-flat scenarios the major contribution to the matter density of the
universe is the term resulting from violations of energy conservation. This
can be seen from 
\begin{equation}
\rho =\frac B{a^{3\gamma }}+\frac{3Kc_0^2\phi n}{4\pi (3\gamma +2n-2)}
a^{2n-2}
\end{equation}
derived above. Taking radiation as an example ($\gamma =4/3$), if $n<-1$ we
are pushed towards the attractor $\Omega =1$ and the energy-violation term
(second term) becomes negligible. In scenarios in which the quasi-flatness
problem is solved the converse happens: the second term dominates. Matter is
continuously being created and this provides the dominant contribution to
cosmic matter at any given time. Matter which was created in the past gets
diluted away by expansion very quickly. In effect, we have a state of
permanent ``reheating''. This scenario has a passing resemblance with the
steady state universe, in which the so called $C$ field continuously creates
more and more matter.

If we couple a changing $c$ say to the Lagrangian of the standard particle
physics model then all particles are produced, and whether they behave like
matter or radiation depends on whether $T\ll m$ or $T\gg m$. If $n<0$ the
energy density in created matter decreases, and the universe cools down as
it expands. Hence there will necessarily be a radiation to matter transition
in these scenarios, just like in the standard Big Bang theory with constant $
G$ and $c$.

A complication arises when we attempt to model the evolution through this
transition because the structure of the attractors changes. For a
radiation-dominated universe one requires $-1<n<0$ for the open attractor $
\Omega =-n$ to be achieved. For a matter-dominated universe the attractors
for open universes are given by $\Omega =-2n$ and are achieved when $
-1/2<n<0 $. Hence, if $-1<n<-1/2$ we have attraction towards a (quasi-flat)
open universe during the radiation-dominated epoch which is then driven
towards flatness in the subsequent matter-dominated epoch. If $-1/2<n<0$ we
have attraction to a (quasi-flat) open model in both epochs but the universe
will evolve from a trajectory asymptoting to $\Omega =-n$ toward one
asymptoting to $\Omega =-2n$ after the matter-radiation equality time. The
solution (\ref{soleps}) is then no longer valid because $\gamma $ is time
dependent.

\section{Solutions with variable $\phi$}

\label{gvar}

\subsection{Exact Radiation solutions}

For radiation, $\gamma =4/3$, we can solve the field equations by
generalising the method introduced for scalar-tensor theories introduced in
ref. \cite{JDBBD}. Define the conformal time

\begin{equation}
ad\tau =dt  \label{time}
\end{equation}
so $^{\prime }=d/d\tau .$ Eq. (\ref{fr3}) integrates to give

\begin{equation}
\phi ^{\prime }=\frac A{a^2}  \label{int}
\end{equation}
with $A$ constant .

Now change the variables of eq.(\ref{fr2}) by introducing

\[
y=\phi a^2 
\]
and $d/d\tau .$ It reduces to

\begin{equation}
y^{\prime 2}=A^2(1+\frac{2\omega }3)+\frac{32\pi y\rho a^4}3-4Kc^2y^2
\label{y1}
\end{equation}
For radiation, eq. (\ref{den1}) gives

\begin{equation}
\rho a^4=B+\frac{3Kc_0^2my^{2m+1}}{4\pi (2m+1)}  \label{y2}
\end{equation}
If we substitute (\ref{y2}) in (\ref{y1}) then we have

\begin{equation}
y^{\prime 2}=A^2(1+\frac{2\omega }3)+\frac{32\pi B\ }3y+4Kc_0^2[\frac{2m}{\
(2m+1)}-1]\ y^{2(m+1)}  \label{y5}
\end{equation}

This can be integrated exactly for appropriate choices of $m$ and its
qualitative behaviour is easy to understand. Note that if we can solve (\ref
{y5}) for $y(\tau )$ $=\phi a^2$, then we know $\phi (\tau ,a)$ and can
solve (\ref{int}) to get $\phi (\tau )$ hence $a(\tau ),$ and finally $a(t)$
from (\ref{time}) since

\begin{equation}
\frac{\phi ^{\prime }}\phi =\frac Ay  \label{y6}
\end{equation}
We note that the curvature term in (\ref{y5}) changes sign for a special
value of $m,$

\[
m=m_{*}=-\frac 12 
\]
Examining (\ref{y5}) we see that the $K=0$ solution appears to be approached
if $2(m+1)<1$, ie $m<-1/2.$

When $m=-1$ there is a simple special case which allows an exact solution.
Eq. (\ref{y5}) is now

\begin{equation}
y^{\prime 2}=\alpha +\beta y+\Gamma  \label{y7}
\end{equation}
where $\alpha ,\beta >0,\Gamma =+4Kc_o^2.$ We integrate (\ref{y6}) to get

\begin{equation}
y=\frac \beta 4(\tau +\tau _0)^2-\frac{\alpha +\Gamma }\beta =\phi a^2
\label{y8}
\end{equation}
Hence,

\[
\phi =\phi _0\left[ \frac{\beta (\tau +\tau _0)-2\sqrt{\alpha +\Gamma }}{
\beta (\tau +\tau _0)+2\sqrt{\alpha +\Gamma }}\right] ^{A/\sqrt{\alpha
+\Gamma }} 
\]
and

\[
a^2=\frac y\phi =\frac{\beta ^2(\tau +\tau _0)^2}{4\beta \phi _0}\left[ 
\frac{\beta (\tau +\tau _0)+2\sqrt{\alpha +\Gamma }}{\beta (\tau +\tau _0)-2
\sqrt{\alpha +\Gamma }}\right] ^{A/\sqrt{\alpha +\Gamma }} 
\]
We see that if $\alpha +\Gamma >0$ then, as $\tau \rightarrow \infty ,$ we
have $a\rightarrow \tau \rightarrow t^{1/2}$ unless the denominator blows up.

If $\alpha +\Gamma <0$ then we get

\[
\ln (\phi /\phi _0)=\frac{2A}{\sqrt{-(\alpha +\Gamma )}}\tan ^{-1}\left( 
\frac{\beta (\tau +\tau _0)}{\sqrt{-(\alpha +\Gamma )}}\right) 
\]
and

\[
a^2=\frac 1{4\phi _0\beta }\left[ \beta ^2(\tau +\tau _0)+\lambda ^2\right]
\exp \{-\frac{4A}\lambda \tan ^{-1}\left( \frac{\beta (\tau +\tau _0)}
\lambda \right) \} 
\]
where

\[
\lambda ^2\equiv -4(\alpha +\Gamma ) 
\]
Looking back at the defining equation (\ref{y5}), we see that for this $m=-1$
case, $\Gamma =4Kc_0^2$ and so

\[
\alpha +\Gamma =A^2(1+\frac{2\omega }3)+4Kc_0^2\ 
\]
and for large $A$ we have $\alpha +\Gamma >0$ when $K>0$.

\subsection{Numerical solutions}
A numerical evolution of the
equations with a varying $\phi $ reveals that for reasonable initial values
of $\phi $ the structure of attractors is not affected. However, the speed
at which attractors are reached may be increased if $\phi $ is allowed to
change. As an example we shall consider the case in which one starts from a
Milne universe ($\epsilon =-1$). In \cite{vsl0} it was argued that this is
the natural initial condition to consider in the context of VSL cosmologies.

\begin{figure}[tbp]
\centerline{{\psfig{file=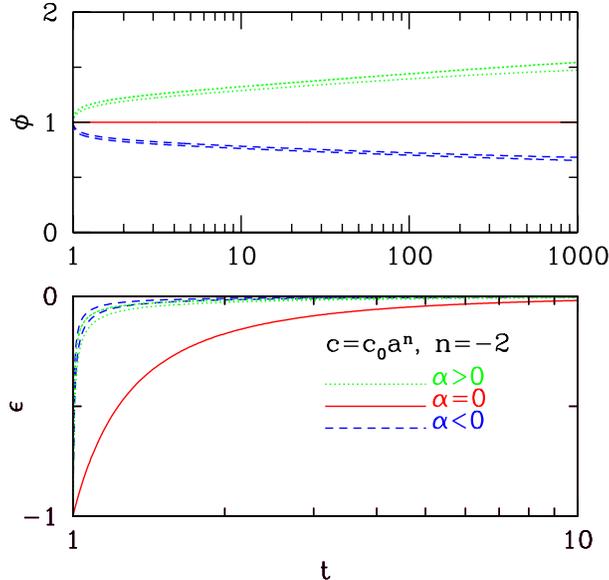,width=8 cm,angle=0}}}
\caption{
The evolution of $\epsilon =\Omega -1$
and of $\phi =1/G$ from a Milne starting point, with $\gamma =4/3$, $
c=c_0/a^2$, and different values of the integration constant $\alpha ,$ of
eq. (\ref{dotG}). We see that flatness is achieved faster with a changing $G$
(increasing or decreasing).}
\label{figbd}
\end{figure}

\begin{figure}[tbp]
\centerline{{\psfig{file=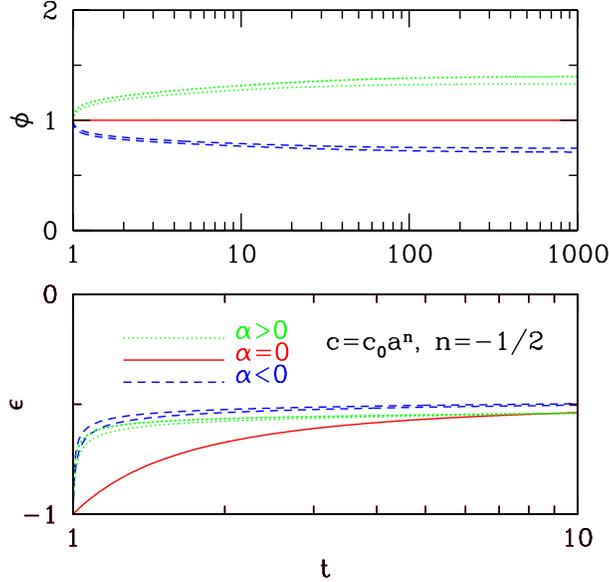,width=8 cm,angle=0}}}
\caption{
The quasi-flat scenario with VSL and a
changing $G$. We consider a Milne starting point, with $\gamma =4/3$, $
c=c_0/a^{1/2}$, and different integration constants $\alpha $. The
quasi-flat attractor is still the same as in the constant $\phi $ case, but
it is achieved faster with a changing $\phi $ (especially a decreasing one).}
\label{figbd1}
\end{figure}

Let us then consider the case $\gamma =4/3$, and integrate Eqns. (\ref{fr1}
), (\ref{fr2}), and (\ref{fr3}). We consider solutions of the form (\ref
{dotG}) with various integration constants $\alpha $. In Figures~\ref{figbd}
and \ref{figbd1} we show the evolution of $\epsilon =\Omega -1$ and $\phi $
from such a state, with $c=c_0a^n$ and $n=-2,-1/2$. In the first case we see
that we still have a flat attractor, which is reached much more rapidly if $
\phi $ is increasing or decreasing. In the latter case we have a quasi-flat
attractor, with $\epsilon =-1/2$, whether or not $\phi $ is allowed to
change. The attractor is achieved faster with a changing $\phi $, especially
a decreasing one.

\section{Asymptotic Solutions to the Flatness Problem}

\label{asym} Exact solutions are only possible for particular equations of
state and $c$-variation laws but it is possible to understand the asymptotic
behaviour in general. The VSL theory Brans-Dicke solutions are given for
eqns. (\ref{fr1})-(\ref{fr3}) together with (\ref{noncons}). We assume a
perfect-fluid equation of state given by (\ref{gamma}). In the flat case ($
K=0$) the equations reduce to those of standard BD flat universes with
constant $c$. At late times, when $K=0$, the general BD solutions for all $
\gamma $ approach the particular ('matter-dominated' or 'Machian') power-law
solutions \cite{nar}

\begin{eqnarray}
G &\propto &\phi ^{-1}\propto t^{-D}  \label{s1} \\
a(t) &\propto &t^A  \label{s2}
\end{eqnarray}
where

\begin{equation}
A=\frac{2+2\omega (2-\gamma )}{
\begin{array}{c}
4+3\omega \gamma (2-\gamma ) \\ 
\end{array}
}  \label{A}
\end{equation}
\begin{equation}
D=\frac{2(4-3\gamma )}{
\begin{array}{c}
4+3\omega \gamma (2-\gamma ) \\ 
\end{array}
}  \label{D}
\end{equation}
Hence, we see that $D+3\gamma A=2;$ note that $D=0$ for radiation;.These
exact power-law solutions reduce the special $\phi =const,a\propto t^{1/2}$
general relativity solution in the radiation case. We shall therefore look
at the stability of these $K=0$ asymptotes when we turn on the $K\neq 0$
terms in eqns. (\ref{fr1})-(\ref{fr3}) and (\ref{noncons}) with $c(t)$
included. If we substitute (\ref{A}) and (\ref{D}) in (\ref{noncons}) with $
c=c_0a^n$ then we have

\begin{eqnarray*}
&& \\
(\rho t^{3\gamma A})^{\cdot } &\simeq &\frac{3Kc_0^2An\phi _0}{4\pi }
t^{A(3\gamma +2n-3)+A-1+D}\simeq \frac{3Kc_0^2An\phi _0}{4\pi }t^{1+2A(n-1)}
\end{eqnarray*}
Integrating, we get

\begin{equation}
\rho t^{3\gamma A}\simeq B+\frac{3Kc_0^2An\phi _0}{8\pi (1+An-A)}
t^{2(1+An-A)}  \label{asy}
\end{equation}
if $A(n-1)\neq -1$. Thus, to solve the flatness problem we will need the $B$
term to dominate the $K$ term on the right-hand side of eq. (\ref{asy}) at
large $t;$ that is, since $A>0$ for expanding universes, asymptotic approach
to flatness requires

\[
n<1-\frac 1A. 
\]
Using (\ref{A}), this gives the condition

\[
n<\frac{(2-\gamma )(2-3\gamma )\omega -2}{2[1+\omega (2-\gamma )]}. 
\]

For radiation and dust universes this condition for solving the flatness
problem reduces to:

\begin{eqnarray}
rad &:&n<-1  \label{co1} \\
dust &:&n<\frac{-(\omega +2)}{2(\omega +1)}\rightarrow -\frac 12\ as{\rm {\ }
\omega \rightarrow \infty}  \label{co2}
\end{eqnarray}
where (\ref{co2}) approaches the general relativity value for $\omega
\rightarrow \infty ,$ as expected and (\ref{co1}) agrees with the radiation
case studied above.

\section{Exact solutions to the flatness problem}

\label{exact}

\subsection{Radiation-dominated case}

The exact solutions found for the VSL theory with $c=c_0a^n$ in \cite{vsl1}
can be generalized to the BD case if one assumes that the variation of the
speed of light is governed by a relation of the form 
\begin{equation}
c=c_0(a\sqrt{G\phi })^n
\end{equation}
This reduces to the previously studied (non-BD) case when $\phi $ is
constant. In the Einstein frame one then has $c=c_0{\hat a}^n$ and in the
radiation-dominated epoch the first of equations (\ref{conseins}) becomes 
\begin{equation}
{\hat \rho }^{\prime }+3{\frac{{\hat a}^{\prime }}{{\hat a}}}({\hat \rho }+{
\hat p}/c^2)={\frac{3Kc^2}{4\pi G{\hat a}^2}}{\frac{c^{\prime }}c}
\end{equation}
Hence, for radiation, one has the integral 
\begin{equation}
{\hat \rho }{\hat a}^4=B+{\frac{3Kc_0^2n{\hat a}^{2n+2}}{4\pi G(2n+2)}}
\end{equation}
with $B$ constant. Since ${\hat \rho }{\hat a}^4=\rho a^4,$ in the Jordan
frame, one has 
\begin{equation}
\rho a^4=B+{\frac{3Kc_0^2na^{2n+2}(G\phi )^{n+1}}{4\pi G(2n+2)}}
\end{equation}
This solution can be verified directly from eq. (\ref{noncons}), although it
would have been difficult to guess it without a foray into the Einstein
frame.

Conditions for solving the flatness problem can now be derived by inspection
of the Friedmann-like equation: 
\begin{equation}
{\left( \frac{\dot a}a\right) }^2={\frac{C_1}{\phi a^4}}+{C_2Ka^{2n-2}\phi ^n
}-{KC_3a^{2n-2}\phi ^n}-{\frac{\dot \phi a}{\phi a}}+{\frac \omega 6}{\left( 
\frac{\dot \phi }\phi \right) }^2
\end{equation}
Since $\phi $ approaches an asymptotic value we still require that $n<-1$
for the term in $C_1$ to dominate the curvature ($K$) term at late times.

If $\phi $ is decreasing we do not need the speed of light to decrease in
time so fast in order to solve the flatness problem. Since 
\begin{equation}
{\frac{\dot c}c}=n{{\frac{\dot a}a}+{\frac 12}{\frac{\dot \phi }\phi }}
\end{equation}
we see that while ${\dot \phi /\phi }$ is non-negligible we have $|\dot
c/c|<|n|\dot a/a$. Weaker gravity in the early universe therefore assists a
varying speed of light in solving the flatness problem

\subsection{Other equations of state}

Introduce the variable

\begin{equation}
x=\phi a^{3\gamma -2\ }  \label{defx}
\end{equation}
and assume that $c$ varies as

\begin{equation}
c=c_0x^{n/2}  \label{cx}
\end{equation}
Hence, (\ref{cons2}) becomes

\[
(\rho a^{3\gamma })^{\cdot }=\frac{3Kc_0^2n}{8\pi }x^n\dot x 
\]
Integrating, we have

\begin{equation}
\rho a^{3\gamma }=B+\frac{3Kc_0^2n\phi ^{n+1}a^{(3\gamma -2)(n+1)}}{8\pi
(n+1)}  \label{den1}
\end{equation}
if $n\neq -1$. This solution can be used to study the evolution for general $
\gamma .$

\section{Solutions to the Quasi-flatness Problem}

\label{qflat}

\subsection{The constant $\phi $ case}

We want to discover if it is ever natural to have evolution which asymptotes
to a state of expansion with a {\it non-critical} density (for example, say, 
$\Omega _0\simeq 0.2,$ as some observations have implied). For simplicity we
consider first the solutions with constant $\phi .$ The conservation
equation (\ref{noncons}) with constant $\phi $ reduces to

\begin{eqnarray*}
\frac{(\rho a^{3\gamma })^{\cdot }}{a^{3\gamma }} &=&\frac{3Kc_0^2n}{4\pi }
a^{2n-3}\dot a\phi \\
&&\ 
\end{eqnarray*}
which integrates to give

\begin{equation}
\rho =\frac B{a^{3\gamma }}+\frac{3Kc_0^2\phi n}{4\pi (3\gamma +2n-2)}
a^{2n-2}  \label{exa}
\end{equation}
with $B$ constant, so substituting in (\ref{fr2}) we have

\[
\frac{\dot a^2}{a^2}=\frac{8\pi B}{3\phi a^{3\gamma }}-\frac{K(3\gamma
-2)c_0^2\ }{a^{2(1-n)}(3\gamma +2n-2)}. 
\]
>From this we can easily determine the attractors at large $a$. Specialising
to the radiation case, we have

\[
\frac{\dot a^2}{a^2}=\frac{8\pi B}{3\phi a^4}-\frac{K\ c_0^2\ }{
a^{2(1-n)}(1+n)}. 
\]

Now, the density parameter $\Omega $ is defined by: 
\begin{equation}
{\frac \Omega {\Omega -1}}={\frac{\frac{8\pi G\rho }3}{\frac{Kc^2(t)\ }{a^2}}
}
\end{equation}
For a quasi-flat open universe the ratio between the two terms on the
right-hand side is approximately constant. From the solution (\ref{exa})
with $\gamma =4/3$ we have 
\begin{equation}
{\frac \Omega {\Omega -1}}={\frac{\frac{8\pi G}3{\left( {\frac B{a^4}}+{
\frac{3Kc_0^2n}{8\pi G(n+1)a^{2(1-n)}}}\right) }}{\frac{Kc_0^2}{a^{2(1-n)}}}}
\end{equation}
If $n<-1$ the energy production term (second term on the right-hand side of (
\ref{exa})) is subdominant at late times, and so this ratio goes to
infinite, meaning $\Omega =1$. If $0<n<-1$ then the second term in (\ref{exa}
) dominates at late times and therefore the expansion asymptotes to one
displaying 
\begin{equation}
{\frac \Omega {\Omega -1}}={\frac n{n+1}}
\end{equation}
that is $\Omega =-n$. So the key feature is that in these scenarios the
curvature terms associated with violations of energy conservation dump
energy into the universe at the same rate as the curvature term in Friedmann
equation. Therefore the ratio of the two terms in the Friedmann equation
stays constant, leading to an open universe with finite $\Omega $ value
today.

\subsection{ The varying $\phi $ case}

The formulation of the radiation case in terms of the variable $y(\tau
)=\phi a^2$ allows us to extend the analysis to the varying-$\phi $ case.
Taking

\begin{equation}
c=c_0y^m  
\end{equation}
we have

\begin{equation}
y^{\prime 2}=A^2(1+\frac{2\omega }3)+\frac{32\pi y\rho a^4}3-4Kc^2y^2
\end{equation}
and

\begin{equation}
\rho a^4=B+\frac{3Kc_0^2my^{2m+1}}{4\pi (2m+1)}.  
\end{equation}
We introduce the density parameter

\[
\Omega =\frac \rho {\rho _c}=\frac{y^{\prime 2}-A^2(1+\frac{2\omega }
3)+4Kc^2y^2}{y^{\prime 2}-A^2(1+\frac{2\omega }3)\ }, 
\]
so that

\[
\frac \Omega {\Omega -1}=\frac{8\pi }{3Kc_0^2y^{2m+1}}\left[ B+\frac{
3Kmc_0^2y^{2m+1}}{4\pi (2m+1)}\right] . 
\]
As $t\rightarrow \infty ,\tau \rightarrow \infty ,y\rightarrow \infty $ we
see that $\Omega \rightarrow 1$ if $2m+1<0$ (in Section {\em \ref{gvar} }we
give a full exact solution for the case $m=-1$ which falls into this class),
but if $2m+1>0$ we have approach to a quasi-flat open universe with

\[
\Omega \rightarrow -2m 
\]
Again, we can have a natural quasi-flat asymptote with $0<\Omega <1.$

\subsection{General asymptotic behaviour}

Consider the behaviour of eq. (\ref{y5}) at large $\tau $ and $y$ in the
case where $2m+1>0;$ that is, where we have the quasi-flat attractor $\Omega
\rightarrow \Omega _\infty =-2m$ as $\tau \rightarrow \infty .$ This assumes
the constants in the Friedmann equation allow sufficient expansion to occur
(so there is no collapse at a finite early time). Since

\[
y^{\prime }\simeq \Gamma y^{m+1} 
\]
we have

\[
y\simeq \left[ \frac{\Gamma \Omega _\infty }2(\tau +\tau _0)\right] ^{\frac
2{\Omega _\infty }}\sim \tau ^{\frac 2{\Omega _\infty }} 
\]
as $\tau \rightarrow \infty .$ Using (\ref{y6}) we get

\[
\phi =\phi _0\exp \left[ \frac{\tau ^{1-\frac 2{\Omega _\infty }}}{1-\frac
2{\Omega _\infty }}\right] 
\]
so, since $\Omega _\infty <1$, we have $\phi \rightarrow \phi _0$ as $\tau
\rightarrow \infty .$ So, we have

\[
a(\tau )\sim \tau ^{\frac 1{\Omega _\infty }}\exp \left[ \frac{\tau
^{1-\frac 2{\Omega _\infty }}}{2(1-\frac 2{\Omega _\infty })}\right] \sim
\tau ^{\frac 1{\Omega _\infty }} 
\]
Thus $t\sim \tau ^{(\Omega _\infty +1)/\Omega _\infty }$ and

\[
a\sim t^{\frac 1{(1+\Omega _\infty )}} 
\]
as $t\rightarrow \infty .$ When $\Omega _\infty =1$ we have the expected $
a\sim t^{\frac 12}$ flat radiation asymptote.

\section{The Lambda and the Quasi-lambda Problems}

\label{qlamb}

\subsection{The General Relativity Case}

Let us consider the impact of a varying speed of light on the general
relativity case. Similar results will occur in the BD case (to which it
reduces exactly in the case of radiation with constant $\phi $). If we wish
to incorporate a cosmological constant term, $\Lambda $, (which we shall
assume to be constant) then we can define a vacuum stress obeying an
equation of state

\begin{equation}
p_\Lambda =-\rho _\Lambda c^2,  \label{vac}
\end{equation}
where

\begin{equation}
\rho _\Lambda =\frac{\Lambda c^2}{8\pi G}.  
\end{equation}
Then, since $G$ is constant, and replacing $\rho $ by $\rho +\rho _\Lambda $
in (\ref{noncons}), we have the generalisation

\begin{equation}
\dot \rho +3\frac{\dot a}a(\rho +\frac p{c^2})+\dot \rho _\Lambda =\frac{
3Kc\dot c}{4\pi Ga^2}.  \label{cons2}
\end{equation}

We shall assume that the matter obeys an equation of state of the form (\ref
{gamma}) and the Friedmann equation is

\begin{eqnarray}
\frac{\dot a^2}{a^2} &=&\frac{8\pi G\rho }3-\frac{Kc^2\ }{a^2}+\frac{\Lambda
c^2}3.  \label{fr} \\
&&\ \   \nonumber  
\end{eqnarray}
We also assume that $c=c_0a^n$ again, so eq. (\ref{cons2}) integrates
immediately to give \cite{vsl1}

\[
\rho =\frac B{a^{3\gamma }}+\frac{3Kc_0^2na^{2(n-1)}}{4\pi G(2n-2+3\gamma )}-
\frac{\Lambda nc_0^2a^{2n}}{4\pi G(2n+3\gamma )}, 
\]
with $B$ a positive integration constant. Substituting in (\ref{fr}) we have

\begin{equation}
\frac{\dot a^2}{a^2}=\frac{8\pi GB}{3a^{3\gamma }}+\frac{
Kc_0^2a^{2(n-1)}(2-3\gamma )}{(2n-2+3\gamma )}+\frac{\Lambda \gamma
c_0^2a^{2n}}{(3\gamma +2n)}  \label{frnew}
\end{equation}
Eq. (\ref{frnew}) allows us to determine what happens at large $a.$

If $-3\gamma >2n>2n-2$ then we see that the flatness and lambda problems are
both solved as before. There are three distinct cases:

\subsubsection{Case 1: $2n>-3\gamma >2n-2$}

The $\Lambda $ term dominates the right-hand side of (\ref{frnew}), the
curvature term becomes negligible, and

\begin{equation}
\frac{\dot a^2}{a^2}\ \rightarrow \frac{\Lambda \gamma c_0^2a^{2n}}{(3\gamma
+2n)}.  \label{lamas}
\end{equation}
So, at large $t,$we have

\[
a\sim t^{\frac{-1}n}. 
\]
Note that for radiation this case requires

\[
-1<n_{rad}<-2, 
\]
while for dust it requires

\[
-\frac 12<n_{dust}<-\frac 32. 
\]
For general fluids it requires

\[
-\frac{\ 3\gamma }2<n<\frac{2-3\gamma }2. 
\]

\subsubsection{Case 2: $2n>2n-2>-3\gamma $}

The scale factor approaches (\ref{lamas}) but the curvature term dominates
the matter density term. Define

\[
\Omega =\Omega _m+\Omega _\Lambda =\frac{8\pi G(\rho +\rho _\Lambda )a^2}{
3Kc^2} 
\]
and then we have that, at large $a$,

\[
\frac \Omega {\Omega -1}\rightarrow \frac{8\pi GB}{3Kc_0^2}a^{2-2n-3\gamma
}. 
\]
Thus $\Omega /(\Omega -1)$ $\rightarrow \infty $ as $a\rightarrow \infty $
for $-3\gamma <-2n+2.$ But when $2n>-3\gamma ,$

\[
\frac \Omega {\Omega -1}\rightarrow \frac{\gamma \Lambda a^2}{K(2n+3\gamma )}
\rightarrow \infty 
\]
in (\ref{frnew}), and so $\Omega \rightarrow 1$.

If $\Lambda =0$ we note that

\[
\frac \Omega {\Omega -1}\rightarrow \frac{2n}{2n-2+3\gamma } 
\]
when $-3\gamma <2n-2$ and this is just the solution for the quasi-flatness
problem found above for general relativity and Brans-Dicke theory in Section 
\ref{qflat}.

For $\Lambda \neq 0,$ when the $\Lambda $ term dominates at large $a$ we see
that

\[
\frac{\Omega _m}{\Omega _\Lambda }=\frac \rho {\rho _\Lambda }\rightarrow -
\frac{2n}{2n+3\gamma } 
\]
and we have a 'solution' to the quasi-lambda problem ({\it ie} the problem
of why $\Omega _m$ and $\Omega _\Lambda $ are of similar order today).
Recall that in this case we have $n<0$, $2n+3\gamma >0$ and so the asymptote
is again of the form $a\sim t^{\frac{-1}n}.$

\section{Challenges for quasi-flat and quasi-lambda scenarios}

Scenarios in which the quasi-flatness problem is solved are considerably
more exotic than the VSL solution to the flatness problem. Unlike flat
scenarios they have a Planck epoch, something which may be a problem. We
discuss this issue in the Appendix.

These scenarios possess several other unusual features. In standard flatness
VSL scenarios, the expansion factor in the radiation-dominated phase is
still $a\propto t^{1/2}$. Standard nucleosynthesis should still be valid
unless there are changes to other aspects of relevant strong and weak
interaction physics (which seems likely). However in scenarios which solve
quasi-flatness we have $a\propto t^{\frac 1{1-n}}$, which for a $\Omega =0.2$
attractor means $a\propto t^{0.83}$. This could easily conflict with the
nucleosynthesis constraints. However it is not enough to state that the
expansion factor at nucleosynthesis time is different: the couplings,
masses, decay times, etc, going into nucleosynthesis are all different \cite
{bek}. One must rework the whole problem from scratch before ruling out
these scenarios on grounds of discordant nucleosynthesis predictions.

Structure formation is another concern. It was shown in \cite{vsl0} that the
comoving density contrast $\Delta $ and gauge-invariant velocity $v$ are
subject to the equations: 
\begin{eqnarray}
\Delta ^{\prime }-{\left( 3(\gamma -1){\frac{a^{\prime }}a}+{\frac{c^{\prime
}}c}\right) }\Delta &=&-\gamma kv-2{\frac{a^{\prime }}a}(\gamma -1)\Pi _T
\label{delcdotm} \\
v^{\prime }+{\left( {\frac{a^{\prime }}a}-2{\frac{c^{\prime }}c}\right) }v
&=&{\left( {\frac{c_s^2k}\gamma }-{\frac 3{2k}}{\frac{a^{\prime }}a}{\left( {
\frac{a^{\prime }}a}+{\frac{c^{\prime }}c}\right) }\right) }\Delta  \nonumber
\\
+{\frac{kc^2(\gamma -1)}\gamma }\Gamma - &&\ kc(\gamma -1)\left( \frac
2{3\gamma }+\frac 3{k^2c^2}\left( \frac{a^{\prime }}a\right) ^2\right) \Pi _T
\label{vcdotm}
\end{eqnarray}
where $k$ is the comoving wave vector of the fluctuations, and $\Gamma $ is
the entropy production rate, $\Pi _T$ the anisotropic stress, and the speed
of sound $c_s$ is given by 
\begin{equation}
c_s^2={\frac{p^{\prime }}{\rho ^{\prime }}}=(\gamma -1)c^2{\left( 1-{\frac
2{3\gamma }\ \frac{c^{\prime }}c}{\frac a{a^{\prime }}}\right) }  \label{cs}
\end{equation}
Note that the thermodynamical speed of sound is given by $c_s^2=(\partial
p/\partial \rho )|_S$. Since in standard Big Bang models evolution is
isentropic: $c_s^2=(\partial p/\partial \rho )|_S=\dot p/\dot \rho
=p^{\prime }/\rho ^{\prime }$. When $\dot c\neq 0$ the evolution need not be
isentropic. However, we keep the definition $c_s^2=p^{\prime }/\rho ^{\prime
}$ since this is the definition used in perturbative calculations. One must
however remember that the speed of sound given in (\ref{cs}) is not the
usual thermodynamical quantity. With this definition one has $\delta
p/\delta \rho =p^{\prime }/\rho ^{\prime }$ for adiabatic perturbations;
that is, the ratio between pressure and density fluctuations mimics the
ratio of its background rate of change.

Let us assume a radiation-dominated background. For superhorizon modes ($
ck\eta \gg 1$) there is a power-law solution, $\Delta \propto \eta ^\beta $
with $\beta =2(n+1)$ and $\beta =n-1$. The general solution takes the form: 
\begin{equation}
\Delta =A\eta ^{2(n+1)}+B\eta ^{n-1}
\end{equation}
where $A$ and $B$ are constants in time. For a constant $c$ ($n=0$) this
reduces to the usual $\Delta \propto \eta ^2$ growing mode, and $\Delta
\propto 1/\eta $ decaying mode. If the flatness problem is to be solved, one
must have $n<-1$. If this condition is satisfied there is no growing mode.

This is an expression, in the context of Machian BD scenarios, of the link
between solving the flatness problem and suppressing density fluctuations.
Flatness is imposed in VSL by violations of energy conservation, acting so
as to leave the universe is a state with $\Omega =1$. This process acts
locally, so it also suppresses density fluctuations. Alternatively, we can
see that the approach to flatness everywhere means that inhomogeneous
variations in the spatial curvature must also die away.

Accordingly, we see that in VSL scenarios there is a strong connection
between solving the flatness problem, and predicting a perfectly homogeneous
universe. Quasi-flat scenarios therefore risk not solving the homogeneity
problem. On the other hand there could some mechanism for amplifying thermal
fluctuations to become seeds for the large scale structure of the universe 
\cite{vsl0}. Indeed the equations above should provide a transfer function
converting the thermal white noise spectrum into a tilted or flat spectrum.
If this were the case, then these theories would predict a link between the
spectral tilt and $\Omega $.

Finally, one may wonder whether such $c(t)$ could ever arise in a dynamical
VSL theory. In work in preparation \cite{BMnew} we address this issue, with
the result that indeed if $\psi =log(c/c_0)$ is a scalar field with a Brans
Dicke type of dynamics: 
\begin{equation}
{\ddot \psi }+3{\frac{\dot a}a}\dot \psi =4\pi G\omega \rho  \label{dyn2}
\end{equation}
then one has a Machian solution with an exponent $n$ related to the coupling 
$\omega $ of the theory. Hence there will be a range of coupling values for
which the flatness problem is solved, and a range for which the
quasi-flatness problem is solved. In the latter case, not only can we
predict an open attractor but also the value of $\Omega $ of the universe is
related to a coupling constant.

\section{Conclusion}

We have performed an extensive analysis solutions to Brans-Dicke theories,
with varying $G$, in which the speed of light is also permitted to vary in
time. We have found cosmological scenarios with novel features. For
power-law variations in the velocity of light with the cosmological scale
factor we identified the cases where the flatness problem can be solved.
These generalise the conditions found in earlier investigations of this VSL
theory. Unlike in inflationary universes which solve the flatness problem,
no unusual matter fields are required in the early universe. We have also
identified the cases in which the quasi-flatness problem can be solved; that
is, where there can be asymptotic approach at late times to an open universe
with a density close to that of the critical value. Similarly, we identified
those variations of $c$ which provided solutions of the lambda problem and
the quasi-lambda problem. The possibility of solving the quasi-flatness and
quasi-lambda problems in this way is a genuine novelty of the VSL theory
that distinguishes from the standard inflationary universe scenario. We have
also discussed some problems with the VSL scenario, highlighting in
particular the matter-radiation transition solutions and the role of the
Planck epoch in setting initial conditions (see Appendix for further
details). We have also examined the behaviour of inhomogeneous perturbations
to the homogeneous and isotropic solutions and found the conditions for
density perturbations to grow or decay.

Our investigations reinforce the conclusions of our earlier investigations
of varying-$c$ theories without varying $G$: the scope for obtaining
cosmological models which share a number of appealing properties, which
closely mirror those of the observed universe, suggest that cosmologies with
varying $c$ should be thoroughly explored. It is a challenge to find
observational predictions which would allow future satellite probes of the
microwave background radiation structure to distinguish them from
inflationary universe models (with constant $c$). We hope that this paper
will serve as a further stimulus to undertake those investigations and to
search out new ways of testing the constancy of the traditional constants of
Nature \cite{webb}.

\section*{Acknowledgments}

JM acknowledges financial support from the Royal Society and would like to
thank A. Albrecht and C. Santos for helpful comments. JDB is supported by a
PPARC Senior Fellowship.

\section*{Appendix - Planck time in VSL scenarios}

The Machian VSL scenario, in which $c=c_0a^n$, introduced by Barrow \cite
{vsl1} has significant advantages to the phase transition scenario, in which
the speed of light changes suddenly from $c^{-}$ to $c^{+},$ preferred by
Albrecht and Magueijo \cite{vsl0}. In the phase transition scenario one runs
into the problem of having to decide when to lay down ``natural initial
conditions'' (that is $\epsilon $ and $\epsilon _\Lambda $ of order 1). For
a phase transition occurring at time $t_c\approx t_P^{+}$, the Planck time $
t_P^{-}$ (built from the constants as they were before the transition) is
much smaller than $t_c$. But why should we lay down natural initial
conditions just before the phase transition? If the only scales in the
problem are the ones set by the constants as they were before the
transition, then natural initial conditions should be set at $t_P^{-}$. If
we are to lay down natural initial conditions at $t_P^{-}$ then the universe
goes off the attractor well before the phase transition. A catastrophic
phase precedes the phase transition, in which the universe becomes Milne
(curvature dominated) or de Sitter ($\Lambda $ dominated). Albrecht and
Magueijo noted that this catastrophic phase is not the end of the universe
in VSL scenarios. A varying speed of light would not write off Milne or de
Sitter universes, but would still push them towards an Einstein-de Sitter
universe. The only universes which would be selected out in this process are
the ones with positive curvature, which would end in a Big Crunch, well
before the phase transition.

The Machian scenario does not have this problem, if we are content with
solving the flatness but not the quasi-flatness problem. If we consider the
radiation dominated phase, $c=c_0a^n$ with $n<-1$, and $a\propto t^{1/2}$,
then we have $t/t_P\propto t^{n+1}$. Hence there is no Planck time in these
scenarios: as $t\rightarrow 0$, one has $t/t_P\rightarrow \infty $. As we go
back in time, the universe becomes hotter and hotter, but the Planck
temperature also increases, and is never achieved at any time. The idea of
setting natural initial conditions at Planck time does not make sense in
these scenarios. We have a universe constantly pushed towards an attractor,
which is flat, and has zero cosmological constant.

Scenarios in which the quasi flatness problem is solved do not have this
desirable feature. We find that $t/t_P\propto t^{\frac{1+n}{1-n}}$. Hence we
must have a Planck time in our past in these scenarios.


\begin{thebibliography}{99}
\bibitem{infl}  A.D. Linde, Inflation and Quantum Cosmology, Academic Press
Inc., 1990.

\bibitem{prebb}  G. Veneziano, Phys. Lett. B {\bf 265}, 287 (1991); M.
Gasperini and G. Veneziano, Astropart. Phys. {\bf 1,} 317 (1993).

\bibitem{mof}  J. Moffat, Int. J. of Physics D {\bf 2}, 351 (1993); J.
Moffat, Foundations of Physics, {\bf 23}, 411 (1993) .

\bibitem{mof1}  J. Moffat, astro-ph/9811390.

\bibitem{vsl0}  A. Albrecht and J. Magueijo, Phys. Rev. D {\bf 59}, 000
(1999).

\bibitem{vsl1}  J.D. Barrow, Phys. Rev. D {\bf 59}, 000 (1999).

\bibitem{vsl2}  J.D. Barrow and J. Magueijo, Varying-$\alpha $ theories and
solutions to the cosmological problems, Phys. Lett. B (in press 1999).

\bibitem{quasi}  J. Barrow and J. Magueijo, A Solution of the Quasi-flatness
and Quasi-lambda Problems, {\it Phys. Lett. }B (in press 1999).

\bibitem{hawk}  S.W. Hawking, Phil. Trans. Roy. Soc. A {\bf 310}, 303 (1984)

\bibitem{super}  S. Perlmutter et al, Ap. J. {\bf 483}, 565 (1997); S.
Perlmutter et al (The Supernova Cosmology project), Nature {\bf 391,} 51
(1998); Garnavich, P.M et al, Ap.J. Letters {\bf 493}, L53 (1998); Schmidt,
B.P. 1998 Ap.J. {\bf 507}, 46; Riess, A.G. et al, AJ 116, 1009 (1998)

\bibitem{bek}  J.D. Bekenstein, Phys. Rev. D {\bf 25}, 1527 (1982)

\bibitem{btoo}  J.D. Barrow and C. O'Toole, in preparation.

\bibitem{JDBBD}  J.D. Barrow, Phys. Rev. D{\bf \ 47}, 5329 (1992)

\bibitem{open}  J.R. Gott III, in {\it Inner Space, Outer Space}, E. Kolb et
al (eds.); M. Bucher, A.S. Goldhaber, and N. Turok, Phys. Rev {\bf D52},
3314 (1995).

\bibitem{anth0}  J.D. Barrow and F.J. Tipler, {\it The Anthropic
Cosmological Principle, }Oxford UP, Oxford (1986).

\bibitem{anth}  A. Vilenkin, astro-ph/9805252; N. Turok and S.W. Hawking,
hep-th/9803156; A. Linde, gr-qc/9802038.

\bibitem{nar}  H. Nariai, Prog. Theo. Phys. {\bf 40}, 49 (1968).

\bibitem{BMnew}  J.D. Barrow and J. Magueijo, preprint.

\bibitem{webb}  M.J. Drinkwater, J.K. Webb, J.D. Barrow, and V.V. Flambaum,
Mon. Not. R. astron. Soc. {\bf 298, 457 (1998); }J.K. Webb, V.V. Flambaum,
C.W. Churchill, M.J. Drinkwater, and J.D. Barrow, Phys. Rev. Lett. (1999 in
press).
\end{thebibliography}
\end{document}